\documentclass[aps,prc,superscriptaddress,twoside,twocolumn,nofootinbib,showpacs]{revtex4}
\usepackage{amsmath,amssymb}
\usepackage{mathtools}
\usepackage{bbold}
\usepackage[usenames, dvipsnames]{xcolor}
\usepackage{braket}
\usepackage{graphicx,bm}
\usepackage{slashed}
\usepackage{booktabs}
\usepackage{tensor}

\usepackage{multirow}

\usepackage{mwe}

\usepackage{changes}

\usepackage{enumerate}
\definechangesauthor[name={Yeunhwan}, color=magenta]{YH}

\allowdisplaybreaks

\newcommand{\msun}{M_{\odot}}

\begin{document}

\title{Predicting the moment of inertia of pulsar J0737-3039A from \\ Bayesian modeling of the
nuclear equation of state}
\date{\today}

\author{Yeunhwan \surname{Lim} }
\email{ylim@tamu.edu}
\affiliation{Cyclotron Institute, Texas A\&M University, College Station, TX 77843, USA}

\author{Jeremy W. \surname{Holt} }
\email{holt@physics.tamu.edu}
\affiliation{Cyclotron Institute, Texas A\&M University, College Station, TX 77843, USA}
\affiliation{Department of Physics and Astronomy, Texas A\&M University, College Station, TX 77843, USA}

\author{Robert J. \surname{Stahulak} }
\email{r.stahulak@me.com}
\affiliation{Department of Physics and Astronomy, University of Utah, Salt Lake City, UT 84112, USA}

\begin{abstract}
We investigate neutron star moments of inertia from Bayesian posterior probability distributions 
of the nuclear equation of state that incorporate information from microscopic many-body theory and 
empirical data of finite nuclei. We focus on PSR J0737-3039A and 
predict that for this 1.338\,$M_\odot$ neutron star the moment of inertia lies in the range
$1.04 \times 10^{45}$\,g\,cm$^{2} < I < 1.51 \times 10^{45}$\,g\,cm$^{2}$ at the 95\% credibility
level, while the most probable value for the moment of inertia is $\tilde I = 1.36 \times 10^{45}$\,g\,cm$^{2}$.
Assuming a measurement of the PSR J0737-3039A moment of inertia to 10\% precision, we study the 
implications for neutron star radii and tidal deformabilities. We also determine the crustal 
component of the moment of inertia and find that for typical neutron star masses 
$1.3 M_\odot < M < 1.5 M_\odot$ the crust contributes $1-6$\% of the total moment of inertia, 
below what is needed to explain large pulsar glitches in the scenario of strong neutron entrainment.
\end{abstract}

\pacs{
21.30.-x,	
21.65.Ef,	
}

\maketitle


\section{Introduction}

Neutron star observations are a promising tool \cite{Lattimer536} to infer the properties of matter at 
extraordinarily high densities on the order of several times that of atomic nuclei. Shortly 
after the discovery \cite{burgay03,lyne04} of the double-pulsar system J0737-3039, it was 
suggested \cite{lyne04,lattimer05} that radio timing observations of star A in the 
pair could lead to a measurement of periastron advance sufficiently precise to resolve 
the effects of relativistic spin-orbit coupling \cite{wex95}, which enters at second order in a 
post-Newtonian expansion of the orbital motion. This in turn would place constraints on neutron
star moments of inertia \cite{damour88}, complementary to ongoing
LIGO/VIRGO gravitational wave observations \cite{abbott17,abbott18} for neutron star tidal 
deformabilities and X-ray pulsar timing measurements \cite{watts16} for neutron star radii. All of 
these efforts aim to shed light on the properties of ultra-dense matter, its equation of state, and 
the possible existence of novel phases of matter \cite{baym73,au74,glen82,glen91,thorsson94,glen98,bunta04,weber05,alford05,brown07,brown08,weissenborn11,weissenborn12a,weissenborn12b,lim14,lim15h,lim17h} 
conjectured to exist in the cores of neutron stars.

Several studies \cite{molnvik84,morrison04,bejger05,piekarewicz14,carreau18,landry18} have already 
provided a range of predictions for neutron star moments of inertia from different theoretical descriptions of the 
dense matter equation of state based on (i) nonrelativistic many-body calculations with realistic 
two- and three-nucleon forces, (ii) nonrelativistic Skyrme effective interactions, (iii) relativistic 
mean field models, (iv) self-bound strange quarks, and 
(v) meta-modeling that includes only empirical and microscopic constraints on the phenomenological
parameters entering in the equation of state. While measurement of a neutron star's
moment of inertia to 10\% precision may be sufficient \cite{morrison04} to distinguish among several of
the qualitatively different models above, a more detailed understanding of the correlations among the
moment of inertia, equation of state, neutron star radius, and other bulk neutron star properties 
can be achieved within a Bayesian statistical framework \cite{raithel16,raithel17,carreau18}. 
In previous work \cite{lim18}, we have constructed a model of the dense
matter equation of state based on a Taylor series expansion in powers of the Fermi momentum. 
Bayesian posterior probability distributions for the model parameters were then constructed
using microscopic predictions \cite{coraggio13,coraggio14,sammarruca15,holt17prc} from chiral effective
field theory \cite{weinberg79,entem03,epelbaum09rmp,machleidt11,epelbaum15,entem17} to 
define prior probability distribution functions together with 
empirical data \cite{dutra12,holt18} for finite nuclei to define likelihood functions. As a first application,
we compared the Bayesian modeling of neutron star tidal deformabilities with the first observational
data from GW170817 \cite{abbott17}.

In the present work we employ the same framework to study the probability distributions for
neutron star moments of inertia as a function of mass, focusing on the distribution of values for the 
1.338 $M_\odot$ neutron star J0737-3039A. We then make predictions for the crustal fraction of
the moment of inertia. This quantity is central to the ongoing debate \cite{chamel12,andersson12,chamel13,piekarewicz14,watanabe17,carreau18} whether the superfluid
angular momentum reservoir in a neutron star inner crust is sufficient to produce the largest pulsar
glitches, such as those observed in the Vela pulsar \cite{cordes88}. Previous studies 
\cite{andersson12,chamel13} that included for the first time a treatment of the neutron band structure  
in the inner crust have found that the ratio of the crustal moment of inertia to the total moment of inertia 
may need to be as large as 7-9\% in order to account for observed pulsar glitches. In the present work, 
we find that such large crustal moments of inertia are statistically unlikely in the neutron star mass range 
1.2-1.5\,$M_\odot$. 
Due to competing effects, we find only a minor correlation between the fractional crustal moment of
inertia and the core-crust transition density, in contrast to previous works \cite{link99,fattoyev10} 
that suggested a strong correlation with the transition pressure at the crust-core interface.

The paper is organized as follows. In Section \ref{edf} we describe the model used to construct
the dense matter equation of state and the resulting neutron star composition and structure,
including a consistent treatment of the inner crust. In
Section \ref{moi} we solve the simultaneous equations for hydrostatic equilibrium together with
the additional equation for the neutron star moment of inertia in the slow-rotation approximation.
In Section \ref{res} we present our predictions for the probability distribution of neutron star
moments of inertia as a function of mass together with correlations among the neutron star
radius, tidal deformability, nuclear symmetry energy slope parameter, as well as the core-crust transition
density and pressure. We also compute the crustal fraction of the 
neutron star moment of inertia and the implications for the standard model of pulsar glitches. 
We end with a summary and conclusions.


\section{Parameterized nuclear energy density functional}
\label{edf}
The nuclear equation of state\,(EOS), which relates the
energy density and pressure at a given baryon number density,
is essential for understanding the phenomenology of compact stars. 
Purely microscopic approaches for computing the cold
dense matter equation of state start from realistic nuclear forces fitted to nucleon-nucleon
scattering data, the deuteron binding energy, and also the properties of few-nucleon systems when 
three-body forces are included. Recently there has been much interest 
\cite{hebeler10,hebeler11,gezerlis13,tews13,coraggio14,roggero14,carbone14,drischler14,hagen14,wlazlowski14,wellenhofer15,tews16,drischler16,holt17prc,sammarruca18}
in deriving constraints on the equation of state from chiral effective field theory, a framework for
constructing the nuclear force that allows 
for the quantification of theoretical uncertainties through variations in the low-energy
constants of the theory, the order in the chiral expansion, and the choice of resolution
scale. However, since the typical momentum-space cutoffs used to regularize ultraviolet-divergent 
loop integrals are on the order of $\Lambda \lesssim 600$\,MeV, 
chiral nuclear potentials are not expected to provide a good description of nuclear matter
for densities larger than about twice saturation density $2n_0 = 0.32$\,fm$^{-3}$. For the lowest-cutoff 
chiral potential ($\Lambda = 414$\,MeV) employed in the present work, the breakdown density is expected 
to be even smaller. Nevertheless, we find it useful and informative to compute the dense matter equation of
state up to $n = 0.32$\,fm$^{-3}$ for all chiral potentials. Extensions to higher densities are strongly model 
dependent, but previous works have employed general polytrope extrapolations \cite{kai10,kai13} 
or speed of sound parametrization \cite{tews18}, allowing for the inclusion of phase transitions
and general conformal bounds \cite{bedaque14} for strongly interacting matter at very high energy densities.

In the present work, we employ a minimal model for the nuclear energy density functional
beyond $n > 2 n_0$ in which we fit predictions from chiral effective field theory to a fourth-order power series 
expansion in the Fermi momentum up to $n = 2 n_0$ and then extrapolate this functional without 
modification to larger densities. We therefore do not explore the widest range of high-density equations of
state that could include phase transitions or higher powers of the Fermi momentum.
The present modeling is therefore expected to be most reliable for neutron stars with $M \lesssim 1.4 M_\odot$, 
where the maximum central density is $n_{\rm max} \lesssim 3n_0$ \cite{lim19a}. For densities larger than 
$2n_0$, nucleons begin to overlap and the description in terms of purely hadronic degrees of freedom 
becomes increasingly questionable. The presence of a phase transition from hadronic to quark matter 
generically leads to an immediate softening of the equation of state and a reduction in the neutron star radius (and
therefore also the moment of inertia), however, the fate of the heaviest neutron stars under hadron-quark phase 
transitions is strongly model dependent \cite{alford13,chamel13,han18}.

One of the primary aims of neutron star observations is to search
for indications of novel phases of strongly interacting matter in neutron star cores, and the minimal model
employed in the present study provides a useful baseline scenario without exotic degrees of freedom
or phase transitions. Specifically we write the energy density as 
\begin{equation}\label{eq:fun}
\begin{aligned}
\mathcal{E}(n,\delta) =  \frac{1}{2m}\tau_n + \frac{1}{2m}\tau_p
+ [1-\delta^2] f_s(n) + \delta^2 f_n(n) \,,
\end{aligned}
\end{equation} 
where $\delta = \frac{n_n - n_p}{n}$ is the isospin asymmetry, $\tau_p$ and $\tau_n$ are the
proton and neutron kinetic densities, and $f_s$ and $f_n$ are the potential energy contributions 
for symmetric nuclear matter and pure neutron matter of the form
\begin{equation}
f_s(n) = \sum_{i=0}^3 a_i\, n^{(2+i/3)} \,, 
\quad
f_n(n) = \sum_{i=0}^3 b_i\,n^{(2+i/3)}\,.
\label{eq:fns}
\end{equation}
The coefficients $a_i$ and $b_i$ are fitted to individual symmetric nuclear matter and pure neutron
matter equations of state computed in many-body perturbation theory using chiral nuclear forces 
up to $n=2n_0$. We have found that the values of the expansion parameters do not depend 
strongly depend on the choice of maximal density. For instance, in previous works \cite{lim18,lim19b} we have 
reduced the maximum density of the fitting range to $n=0.25$\,fm$^{-3}$ and found only small quantitative
differences in the fitting parameters and derived probability distributions. We include ten equations of 
state obtained by varying the chiral order of the nucleon-nucleon potential 
from next-to-next-to-leading order (N2LO) to N3LO, the 
order in many-body perturbation theory from second to third order, and finally the momentum-space
cutoff from $\Lambda \simeq 400 - 500$\,MeV. The mean and covariance matrices for $\{a_i\}$ and 
$\{b_i\}$ then define our prior Gaussian probability distributions.

Since none of the adjustable parameters in chiral nuclear potentials are explicitly fitted to the 
properties of nuclear matter, we then incorporate empirical constraints on the nuclear matter equation 
of state into Bayesian likelihood functions involving the $\{a_i\}$ and $\{b_i\}$ parameters. For symmetric nuclear
matter, we consider the saturation 
density $n_0$, saturation energy $B$, incompressibility $K$, and skewness $Q$. The values 
for these quantities are typically obtained from fitting energy density functionals to the binding energies 
and charge radii of finite nuclei. We consider 205 high quality models \cite{dutra12}, which give the 
average and standard deviations 
$n_0 = 0.160 \pm 0.003$\,fm$^{-3}$,
$B =15.939 \pm 0.149$\,MeV,
$K = 232.65 \pm 7.00$\,MeV,
$Q = -373.26 \pm 13.91$\,MeV. The parameters $\{a_i\}$ are then uniquely determined in terms of the
empirical nuclear matter parameters $\{n_0, B, K, Q\}$, which provides a method to derive Bayesian 
likelihood functions involving the $a_i$. Note that we choose 205 Skyrme force models among 240
considered in Ref.\ \cite{dutra12}. Some of these Skyrme models do not reproduce the well known 
nuclear matter properties at saturation density. For instance, Skyrme models for which
$n_0 < 0.15$\,fm$^{-3}$, $B< 15$\,MeV, $B>17$\,MeV, and $K>300$\,MeV have been omitted. 
We note that the statistical uncertainty in some empirical parameters, especially $Q$, come out rather 
small when computed over the 205 Skyrme forces. However, we have doubled the standard deviations
on both $K$ and $Q$ and found only small quantitative changes on the order of $<1\%$ to the neutron
star properties computed below.

\begin{figure}
\centering
\includegraphics[scale=0.50]{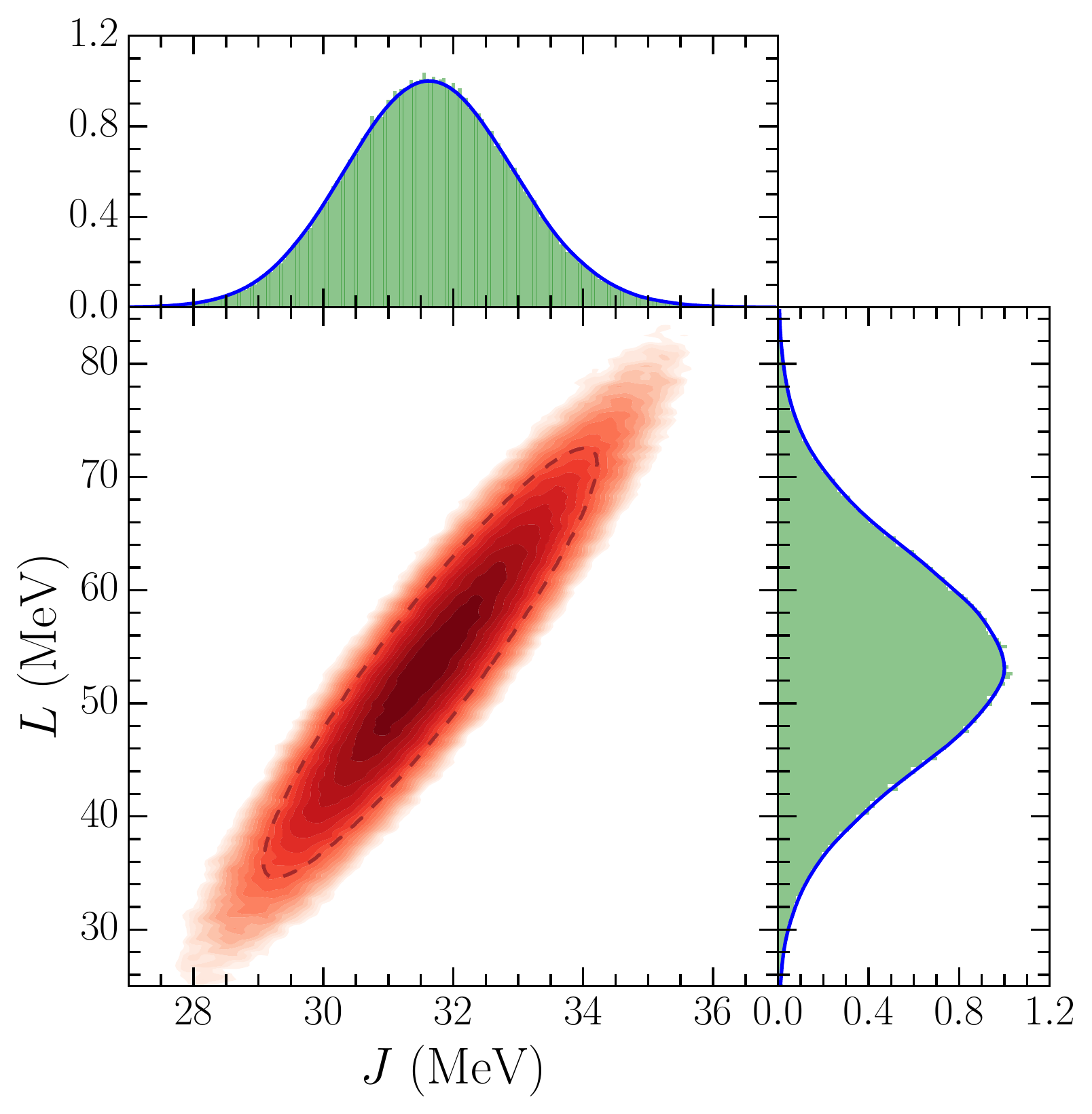}
\caption{Distribution of nuclear symmetry energy and its density slope at the nuclear saturation density from 
	the Bayesian modeling in this work.}
\label{fig:svlcorr}
\end{figure}

For pure neutron matter there are no strong
empirical constraints on the equation of state. However, the nuclear symmetry energy 
$E_{\rm sym} = \frac{E}{A}(n,\delta = 1) - \frac{E}{A}(n,\delta = 0)$ is closely related to the 
isospin asymmetry energy $J = \frac{1}{2}\frac{\partial E/A}{\partial \delta^2} |_{\delta = 0, n=n_0}$, which can be 
extracted from the binding energies of nuclei, giant dipole resonances, neutron skin thicknesses, and 
heavy-ion collision flow data. Recent analyses \cite{Lattimer:2012xj,tews17} give $J = 31 \pm 1.5$\,MeV.
For comparison we note that the range of theoretical predictions for the pure neutron matter equation of state 
at saturation density derived from our chiral EFT calculations is $E/A = 13.8 - 19.7$\,MeV, corresponding to
values of the symmetry energy in the range $29.5\,{\rm MeV} \le E_{\rm sym} \le 36.0\,{\rm MeV}$,
conservatively assuming $B = 16.0 \pm 0.3$\,MeV. In addition, the 205 selected Skyrme forces give the range 
$\langle J \rangle = 31.3$\,\rmfamily{MeV}, $\sigma_{J} = 6.6$\,\rmfamily{MeV}, which is larger than
that obtained from a comprehensive analysis of nuclear experimental data in Refs.\ \cite{Lattimer:2012xj,tews17}.
From correlations \cite{holt18,margueron18} among the isospin asymmetry energy $J$, its slope 
parameter $L$ and curvature $K_{\rm sym}$, likelihood functions involving the $b_i$ were obtained
in Ref.\ \cite{lim18} to derive the final Bayesian posterior probability distributions.
Fig.\ \ref{fig:svlcorr} shows the $J$ and $L$ distribution of the Bayesian modeling in the present work. 
Both $J$ and $L$ have their own Gaussian-like distributions, but there exists a strong correlation between $J$ and $L$
that can be seen also in nuclear mass model calculations \cite{UNEDF0,Lattimer:2012xj}.
In this work, we obtain $29.1\,{\rm MeV} \le J \le 34.4$\,MeV and $34.6\,{\rm MeV} \le L \le 72.6$\,MeV, 
with the correlation $R_{JL}=0.93$ and the slope $\tan{\alpha}=7.94$, from our Bayesian posterior probability distributions.

Randomly sampling from the $a_i$ and $b_i$ joint probability distributions, we construct 300,000
equations of state for nuclear matter in beta equilibrium. For the crust EOS, we utilize a liquid drop model 
with the same nuclear model used in the bulk matter 
equation of state\,\cite{Lim2017a}. This provides
a unified approach, since the structure is constructed
by a single nuclear model, without the need to stitch together various bulk matter EOSs with one specific crust EOS.
Since we use the same nuclear force model for both the core and crust,
the core-crust density is treated consistently, and therefore the moment of inertia
from the crust can be calculated correctly.

In the liquid drop model at $T=0$\,MeV, the total energy density 
is given by\,\cite{LSEOS,lim17}
\begin{equation}
\begin{aligned}
\varepsilon =  & u n_i f_i + \frac{\sigma(x_i)u d}{r_N}
+ 2\pi (n_i x_i e r_N)^2  u f_d(u) \\
& + (1-u)n_{no}f_{no} \,,
\end{aligned}
\end{equation}
where $u$ is the volume fraction of a heavy nucleus in the Wigner-Seitz cell,
$n_i$ is the baryon number density of the heavy nucleus,
$f_i$ is the energy density of the heavy nucleus, 
$\sigma(x_i)$ is the surface tension as a function of
proton fraction\,($x_i$) of the heavy nucleus,
$d$ is the dimension of the nuclear phase (e.g., spherical, cylindrical, or slab),
$r_N$ is the radius of the heavy nucleus,
$f_d(u)$ is the geometric function that reflects the nuclear pasta phase,
$n_{no}$ is the neutron density outside of the heavy nucleus,
and
$f_{no}$ is the energy density of the neutron gas.
The nuclear configuration is determined by minimizing the energy density
for a given total baryon number density $n$ and proton fraction $x$.
In beta equilibrium matter, we include electrons and find
the proton (or electron) fraction $x$ that minimizes the total energy density.
Note that we use the same energy density functional
for $f_{i}$ and $f_{no}$ as used in the bulk matter EOS.
By applying the Lagrange multiplier method with the constraints for
baryon and charge number density, we have\,\cite{lim17}
\begin{equation}
\begin{aligned}
\mu_{ni} -\frac{x_i \sigma^\prime(x_i)d}{r_N n_i} & = \mu_{no}\,, \\
p_i -2\pi (n_i x_i e r_N)^2 \frac{\partial(uf_d)}{\partial u} & = p_{no}\,, \\
n - un_i -(1-u)n_{no} & = 0 \,,\\
nx - u n_i x_i &=0\,.
\end{aligned}
\end{equation}

The core-crust transition density is found by comparing the energy density
of inhomogeneous matter and uniform nuclear matter.
Near the phase boundaries, we employ a Maxwell construction 
to find the exact density and pressure,
\begin{equation}
p_{\text{inh.}} = p_{\text{uni.}} \,, 
\quad \mu_{n_{\text{inh.}}} = \mu_{n_\text{uni.}}
\end{equation}
where `inh.'(uni.) stands for inhomogeneous (uniform) matter.


\section{Moment of inertia}
\label{moi}

The neutron star mass and radius are obtained by solving
the Tolman-Oppenheimer-Volkov (TOV) equation, which is the hydrostatic equilibrium equation for spherically
symmetric neutron stars, given by
\begin{subequations}
\begin{align}
\frac{dp}{dr} & = - \frac{(\varepsilon + p)(m +4\pi r^3 p)}{r(r-2m)} \,, \\
\frac{dm}{dr} & = 4\pi r^2 \varepsilon\,,
\end{align}
\end{subequations}
where $p$ is the pressure, $\varepsilon$ is the energy density (including rest mass), and
$m$ is the enclosed mass within the distance $r$ from the center.
The neutron star moment of inertia is then calculated by solving the
conventional TOV equation with an additional equation including
the rotational frequency.
In a slowly rotating neutron star, the moment of inertia is given by
\cite{hartle67,lattimer2000}
\begin{equation}
I = \frac{8\pi}{3}\int_0^R r^4 (\varepsilon + p)e^{(\lambda -\nu)/2}\,
\frac{\bar\omega}{\Omega} \,dr\, ,
\end{equation}
where $\lambda$ and $\nu$ are metric functions defined by
\begin{eqnarray}
	e^{-\lambda}  & =& \left(1 - \frac{2m}{r}\right)^{-1}\,, \\
\frac{d \nu}{dr}  & = &-\frac{2}{\varepsilon + p} \frac{dp}{dr}\,,
\end{eqnarray}
 $\Omega$ is the angular velocity of a uniformly rotating neutron star,
and $\bar{\omega}$ is the rotational drag function. 
The unitless frequency $\tilde{\omega} = \frac{\bar{\omega}}{\Omega}$
satisfies 
\begin{equation}\label{eq:omegadrag}
\frac{d}{dr} \left(r^4 j \frac{d\tilde{\omega}}{dr} \right)
= - 4r^3 \tilde \omega \frac{d j}{dr}\,,
\end{equation}
where $j=e^{-(\lambda + \nu)/2}$. This rotational drag $\tilde{\omega}$
meets the boundary conditions 
\begin{equation}\label{eq:omgbound}
\tilde{\omega}(r=R) = 1 -\frac{2I}{R^3}\,\quad {\rm and} \quad
\frac{d\tilde{\omega}}{dr}\bigg{\vert}_{r=0} =0\,
\end{equation}
at the surface and center of the neutron star.
The moment of inertia can be integrated from
\begin{equation}\label{eq:moi}
\begin{aligned}
I & =\frac{2}{3}\int_0^R r^3\,\tilde{\omega}  \frac{dj}{dr} \,dr
= \frac{1}{6}\int_0^R \frac{d}{dr} \left(r^4 j \frac{d\tilde{\omega}}{dr}\right)\,dr \\
&= \frac{R^4}{6}\frac{d\tilde{\omega}}{dr}\bigg\vert_{r=R}\,.
\end{aligned}
\end{equation}
The second-order differential equation~\eqref{eq:omegadrag} can be translated to a 
first-order differential equation by introducing
$\phi = d\ln \tilde{\omega} / d \ln r $, giving
\begin{equation}
\begin{aligned}
\frac{d\phi}{dr} & = -\frac{\phi}{r}(\phi + 3) - (4+\phi)\frac{d\ln j}{dr} \\
& =  -\frac{\phi}{r}(\phi + 3) + (4+\phi)\frac{4\pi r^2(\varepsilon + p)}{(r-2m)} \,,
\end{aligned}
\end{equation}
with the boundary condition 
\begin{equation}
\phi(r=0)=0\,.
\end{equation} 
The total moment of inertia of a neutron
star is then given as
\begin{equation}
I =\frac{R^3}{6}\phi_R \tilde{\omega}_R = \frac{\phi_R}{6}(R^3-2I)\,,
\quad
I =\frac{R^3 \phi_R}{ 6 + 2 \phi_{R} } \,,
\end{equation}
with the boundary condition in Eq.~\eqref{eq:omgbound}.

The moment of inertia of the core is given by integrating Eq.~\eqref{eq:moi}
up to the core radius $r=R_t$:
\begin{equation}
I_c = \frac{R_t^4}{6}\frac{d\tilde{\omega}}{dr} \bigg\vert_{r=R_t}
= \frac{R_t^3}{6}\phi_t \tilde{\omega}_t \,.
\end{equation}
From the relation between $\phi$ and $\tilde{\omega}$, we have
\begin{equation}
\tilde{\omega}_t = \tilde{\omega}_R\, \mathrm{Exp} \left[ {-\int_{R_t}^{R} \frac{\phi(r)}{r}}\,dr \right]\,.
\end{equation}
Thus, the moment of inertia of the crust is given as
\begin{equation}\label{eq:icrust}
\begin{aligned}
\Delta I  & = I - I_c = \frac{R^3}{6}\phi_R\tilde{\omega}_R
\left[ 1 - \left( \frac{R_t}{R} \right)^3 \frac{\phi_t}{\phi_R} \frac{\tilde{\omega}_t}{\tilde{\omega}_R} \right] \\
  &
  = I\left\{ 1 - \left( \frac{R_t}{R} \right)^3 \frac{\phi_t}{\phi_R}\, \mathrm{Exp}
      \left[ {-\int_{R_t}^{R} \frac{\phi(r)}{r}}\,dr \right] \right\}\,.
\end{aligned}
\end{equation}
It is known that the slow rotation approximation is valid for J0737-3039A by comparing to
the exact numerical solution without approximation \cite{morrison04}. The error between the exact solution
and slow rotation approximation is estimated by $(\Omega/\Omega_{\rm max})^2$ where $\Omega_{\rm max}
\approx (GM/R^3)^{1/2}$ is the Kepler frequency at the equator of neutron stars. 
For a neutron star with $1.4\,\msun$ with $12\,$km radius,
the Kepler frequency is around $7.9\times 10^3$Hz. Thus, it is expected that for most millisecond pulsars
the slow rotation approximation is valid.


\section{Results}
\label{res}

In the present section we analyze 300,000 equations of state randomly sampled from the Bayesian
posterior distributions for the $a_i$ and $b_i$ parameters entering into the nuclear energy
density functional of Eq.\ (\ref{eq:fun}). For each equation of state we consider up to 110 representative 
neutron stars with masses in the range $1.0 M_\odot \leq M \leq 2.1 M_\odot$ with spacing 
$\Delta M = 0.01 M_\odot$. In the case that the particular equation of state yields a maximum mass 
such that $M_{\rm max} < 2.1 M_\odot$, which occurs for the softest equations of state generated, 
we use the same mass spacing but with a modified range $1.1 M_\odot \leq M \leq M_{\rm max}$.
In total we therefore consider more than 30,000,000 neutron stars for analysis, each constructed with a 
realistic crust equation of state.

\begin{figure}[t]
\centering
\includegraphics[scale=0.48]{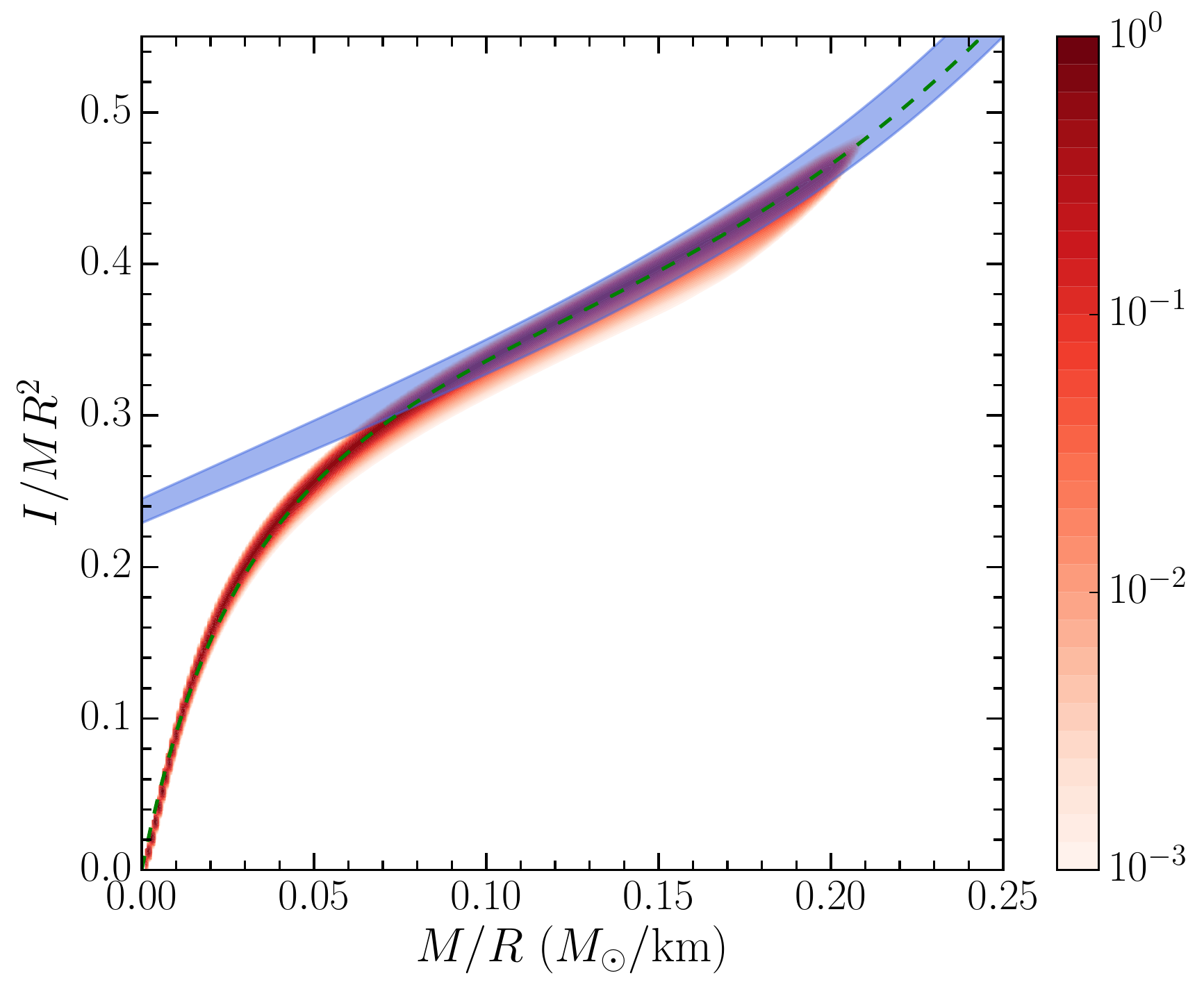}
\caption{(Color online) Probability distribution (red) for the ratio of the neutron star moment of inertia 
$I$ to $MR^2$ as a function of the compactness parameter $M/R$. The distribution is obtained by
randomly sampling 300,000 configurations from the Bayesian posterior probability distributions. The 
empirical (blue) band from Ref.\ \cite{lattimer05} is shown together with the fitting function 
(green dotted) in Eq.\ (\ref{empi}).}
\label{fig:mommr}
\end{figure}

\begin{figure}[t]
	\centering
	\includegraphics[scale=0.48]{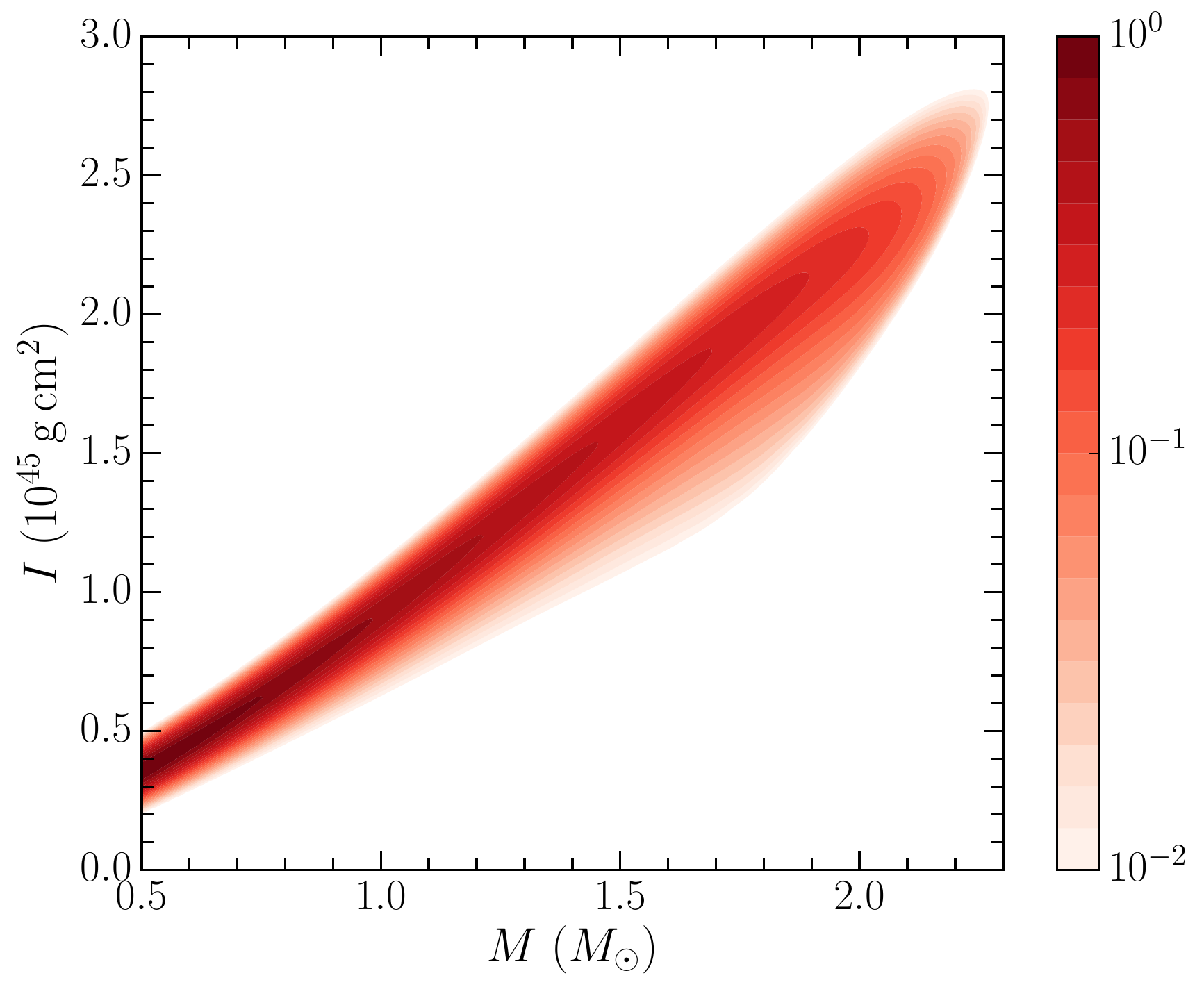}
	\caption{(Color online) Probability distribution for the neutron star moment of inertia as a function of its
	mass from the 300,000 equations of state randomly sampled from the Bayesian posterior distribution.}
	\label{fig:itotm}
\end{figure}

\begin{figure}[t]
	\centering
	\includegraphics[scale=0.48]{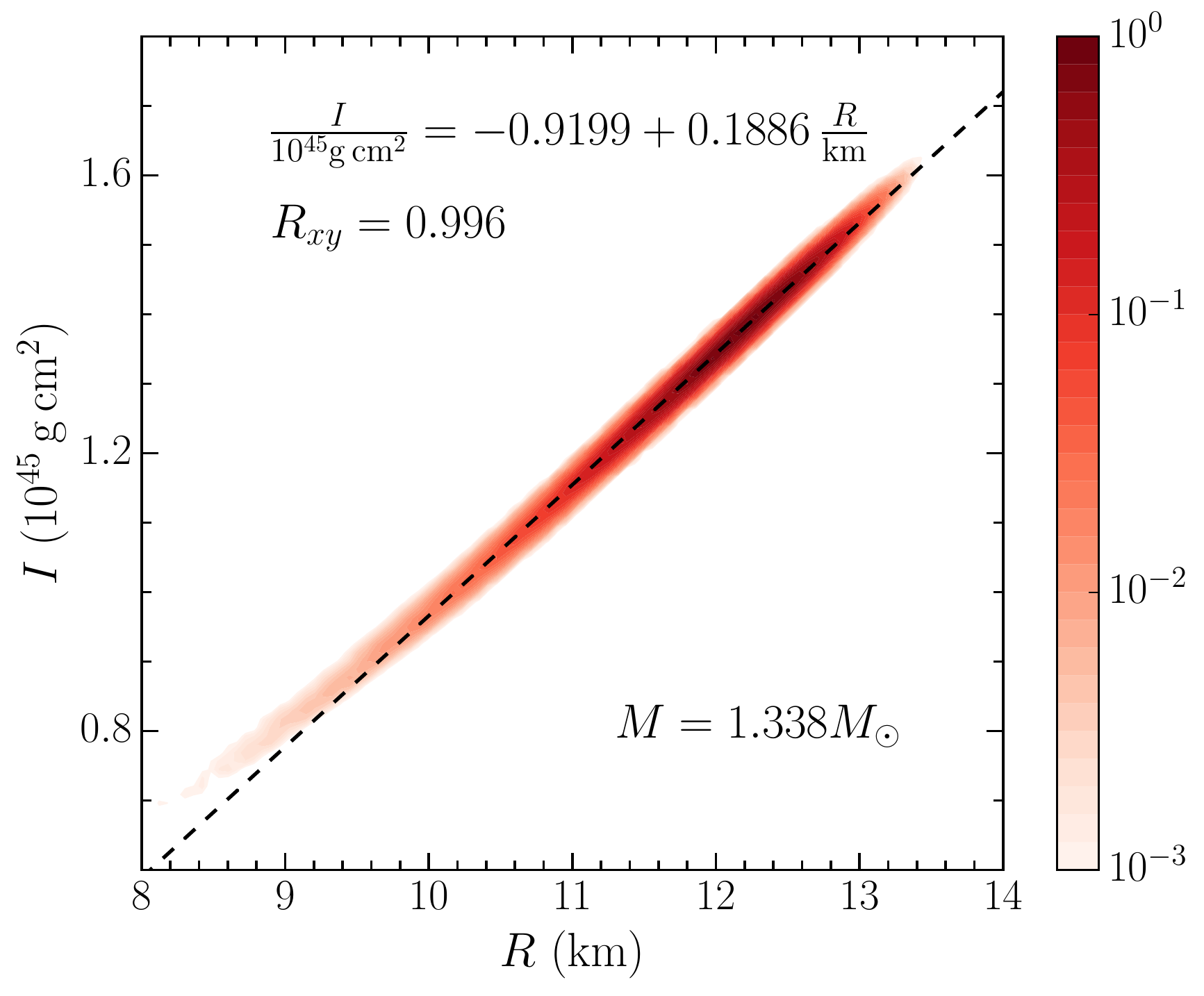}
	\caption{(Color online) Probability distribution (red) for the neutron star moment of inertia 
	vs.\ radius at a fixed mass of $M=1.338\,\msun$ from a Bayesian analysis of the nuclear equation 
	of state. The empirical relation is shown as a black dashed line.}
	\label{fig:moiradius}
\end{figure}

In Fig.\ \ref{fig:mommr} we show in red the resulting probability distribution for the ratio of the moment of inertia 
$I$ to $MR^2$ as a function of $M/R$. In the previous work \cite{lattimer05} it was shown that in the absence 
of phase transitions and other effects that strongly soften the equation of state beyond a few times
normal nuclear densities, there is a nearly
unique relation between the quantity $I/MR^2$ and $M/R$. This relation is shown as the blue band in
Fig.\ \ref{fig:mommr} which we have generated from the empirical formula obtained in Ref.\,\cite{lattimer05}.
Observational evidence suggests that neutron star masses lie in the range
$1.2\,\msun \le M \le 2.0\,\msun$ while radii lie in the range $9\,\mathrm{km} \le R \le 15\,\mathrm{km}$. 
Therefore, we expect only the region
$0.08\,\frac{\msun}{\mathrm{km}}<M/R<0.22\,\frac{\msun}{\mathrm{km}}$ to be physically relevant.
For this range of neutron star compactness parameters $C=M/R$ our results are completely consistent 
with the empirical relation in Ref.\,\cite{lattimer05}. We find that over the entire range of neutron star 
compactnesses, the following formula holds:  
\begin{equation}
	\frac{I}{MR^2} = \frac{M/R + a(M/R)^4 }{b + c (M/R)}\,,
	\label{empi}
\end{equation}
where $a = 27.178\,(\msun/{\rm km})^{-3}$, $b=0.0871\, \msun/{\rm km}$, and $c=2.183$. 
This formula is shown as the green dotted curve in Fig.\ \ref{fig:mommr} and 
should be accurate for most neutron star configurations. 
The relative error for $M/R=0.01\,\msun/{\rm km}$ ($M/R=0.213\,\msun/{\rm km}$) is $3.2\%$ ($1.05\%$).
In the case of $M=1.44\,\msun$, $R=12\,\rm{km}$, which is the canonical average of neutron stars,
the relative error is only $0.06\%$.

In Fig.\ \ref{fig:itotm} we show the neutron star moment of inertia as a function of mass,
plotted as a probability distribution based on the 300,000 equations of state sampled from our Bayesian
posterior distribution. Naturally the moment of inertia increases approximately linearly with the mass.
The uncertainty also generically increases with the neutron star 
mass up to about $M \simeq 1.8 \msun$. Beyond this value, the fraction of equations of state capable of
producing such massive neutron stars decreases as does the range of allowed radii for a given mass. This
results in a narrowing of the moment of inertia probability distribution for the largest-mass neutron stars
in our sample.

\begin{figure}[t]
	\centering
	\includegraphics[scale=0.45]{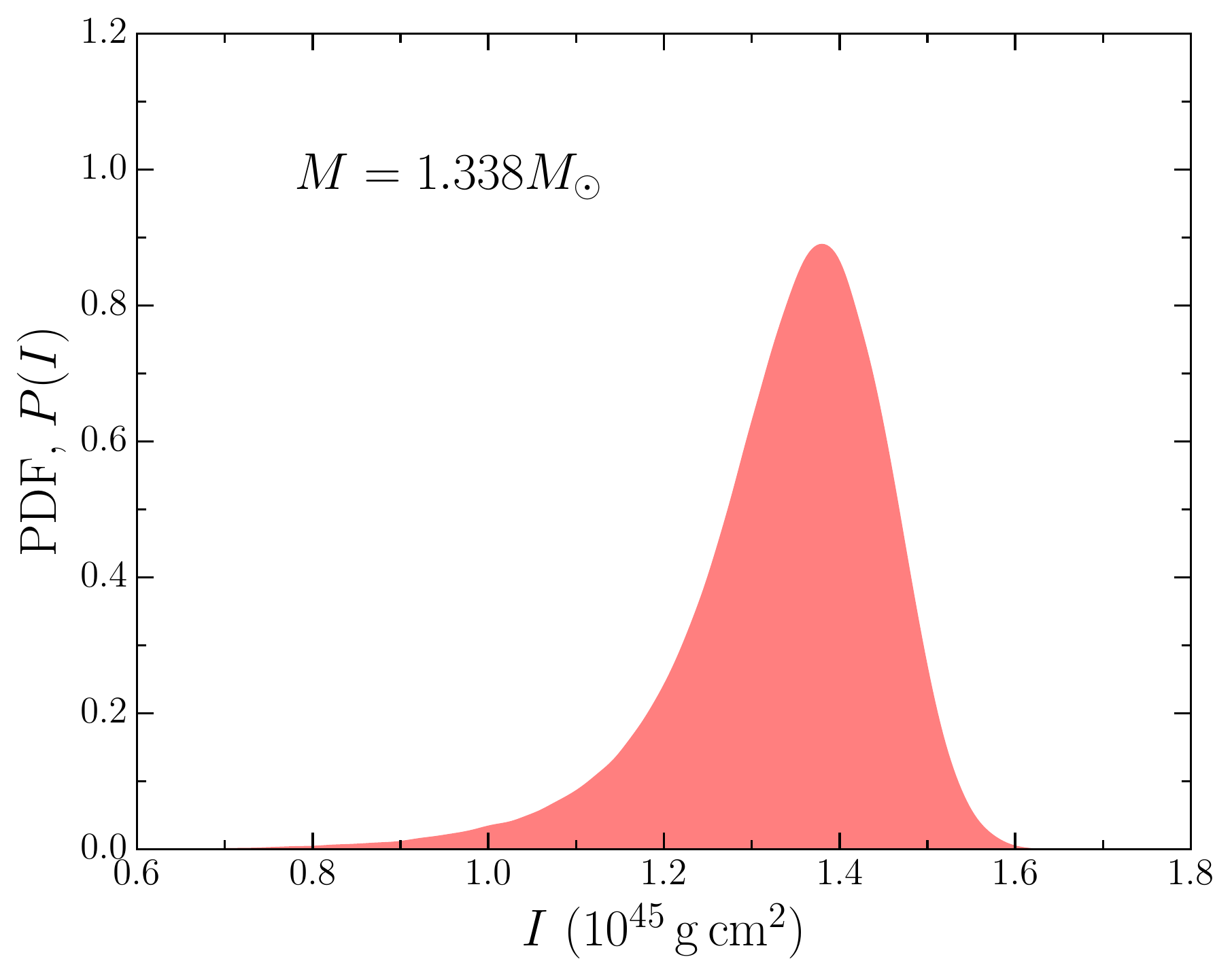}
	\caption{(Color online) Probability distribution for the moment of inertia of a 1.338$\msun$ neutron star
	based on the energy density functionals constrained by nuclear theory and experiment.}
	\label{fig:m1338moidist}
\end{figure}

From Fig.\ \ref{fig:mommr} we see that a simultaneous measurement of neutron star mass and moment of inertia 
will indeed strongly constrain the radius. This is demonstrated more explicitly in Fig.\
\ref{fig:moiradius}, where we plot the probability distribution for the neutron star moment of inertia
vs.\ neutron star radius for a fixed mass of $M=1.338 \msun$ corresponding to that of PSR J0737-3039A.
A total of 300,000 samples are considered, one for each of the generated equations of state. 
We see that the moment of inertia lies in the range 
$1.04 \leq I_{45} \leq 1.51 $\,(95\% credibility), where
for convenience we have defined $I \equiv I_{45} \times 10^{45}{\, \rm g\, cm^2}$, while
the radius lies between $10.3\,{\rm km} \leq R \leq 12.9\,{\rm km}$\,(95\% credibility). In addition we 
observe an approximate linear correlation between the moment of inertia and the radius in this regime 
of the form
\begin{equation}
I = \left (-0.9199 + 0.1886 \frac{R}{\rm km} \right ) \times 10^{45}{\, \rm g\, cm^2}
\end{equation}
with correlation coefficient $R=0.996$. 
From Fig.\ \ref{fig:moiradius} we observe that the moment of inertia probability distribution is asymmetric 
and peaks around $I \simeq 1.36 \times 10^{45}{\, \rm g\, cm^2}$ and $R \simeq 12.2$\,km. We present in
Fig.\ \ref{fig:m1338moidist} the full probability distribution function for the moment of inertia of a $1.338 \msun$
neutron star from the 300,000 EOSs constructed in this work. The probability function appears as an 
asymmetric Gaussian function and therefore the average value of the total moment of 
inertia $\langle I \rangle$ does not match the most probable moment of inertia, which we denote by $\tilde I$.
In this work, we find $\langle I \rangle = 1.338\times 10^{45}\,\mathrm{g\,cm}^2$ with one- and two-sigma 
credibility ranges
given by $I_{-\sigma}=1.233$, $I_{+\sigma} = 1.443$, $I_{-2\sigma} = 1.035$, and  $I_{+2\sigma}=1.514$
in units of $10^{45}\,\mathrm{g\,cm}^2$. In comparison the most likely value of the moment of inertia is
found to be $\tilde I=1.355\times 10^{45}\,\mathrm{g\,cm}^2$.

\begin{figure}[t]
	\centering
	\includegraphics[scale=0.27]{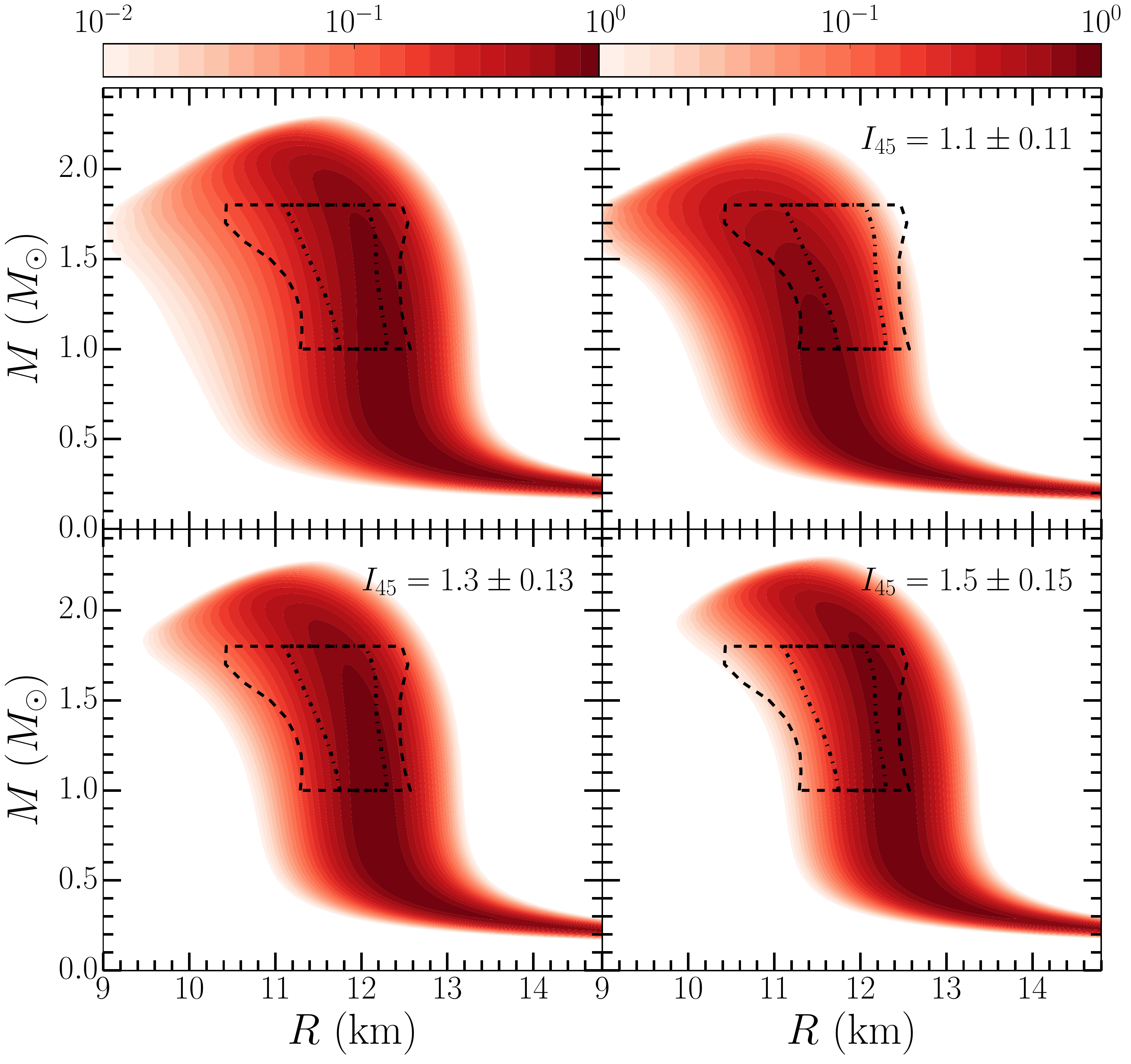}
	\caption{(Color online) Posterior probability distributions for the neutron star mass vs.\ radius relation from 
	artificial moment of inertia measurements with 10\% uncertainties. }
	\label{fig:bayemrdist}
\end{figure}

A precise moment of inertia measurement to 10\% precision is expected to
be competitive with gravitational wave constraints on neutron star radii \cite{abbott18} as well as direct measurements
of radii from X-ray observations. We now discuss the implications for such a 
moment of inertia measurement and how it can be implemented in the current Bayesian modeling
of the equation of state.
In the top left panel of Fig.\ \ref{fig:bayemrdist} we show the neutron star mass and radius distribution 
resulting from our Bayesian analysis of the nuclear energy density functional including constraints from
microscopic many-body theory and nuclear experiments. The softest equations of
state generated from the statistical sampling do not reach a maximum neutron star mass of
$M_{\rm max} = 2\msun$, but this is due primarily to our smooth continuation of the equation of state
beyond twice saturation density. In particular we cannot rule out the existence of higher-power repulsive 
contributions to the nuclear energy density functional beyond $n>2n_0$ that might sufficiently stiffen the
equations of state to support a $2\msun$ neutron star. In the present work we do not focus on the 
properties of the heaviest neutron stars and therefore we keep the softest equations of state in our 
subsequent analysis.

\begin{figure}[t]
	\centering
	\includegraphics[scale=0.48]{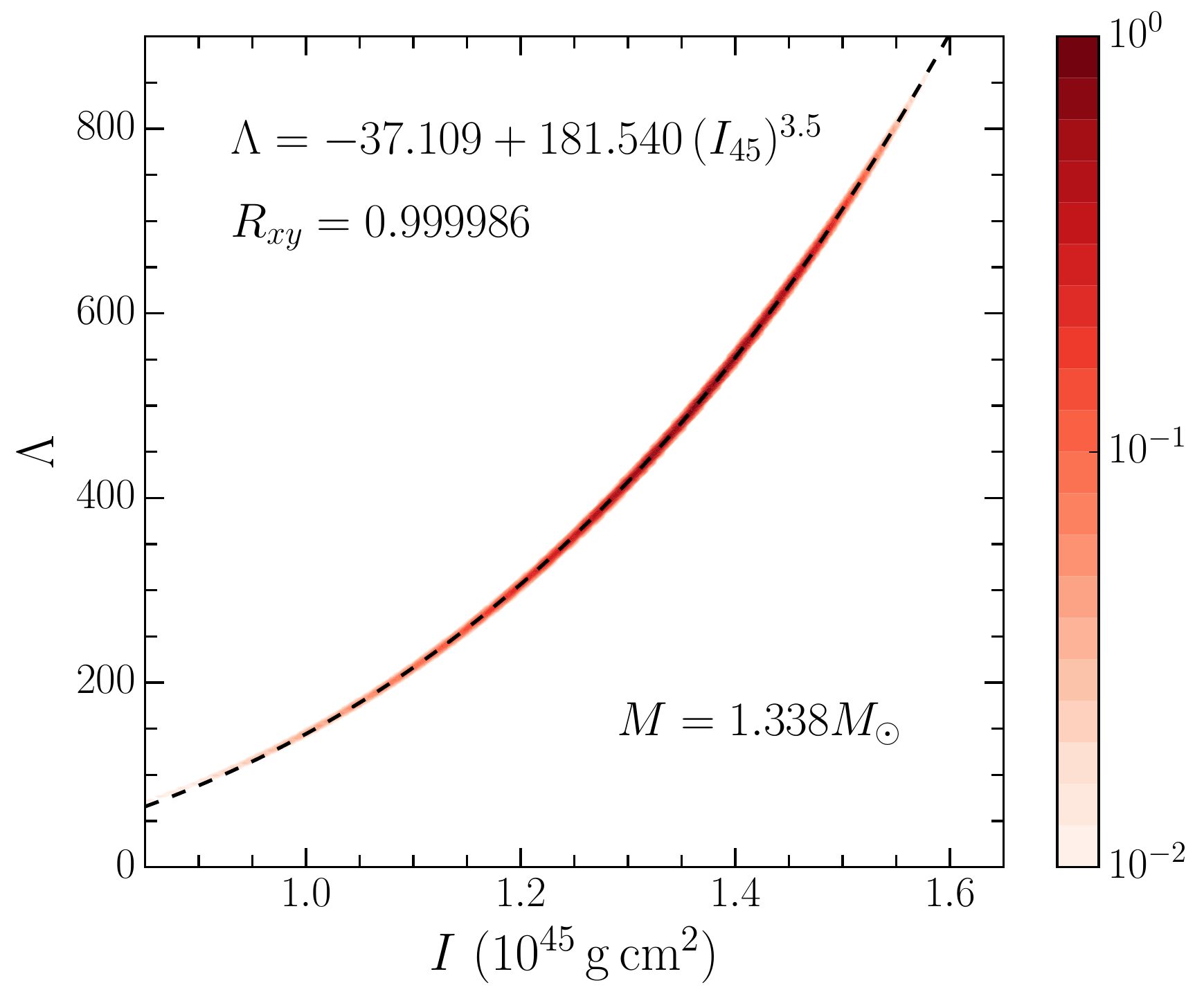}
	\caption{(Color online) Probability distribution for the tidal deformability vs.\ moment of inertia for a neutron star
	with $M=1.338\,\msun$ obtained from the Bayesian posterior probability distribution constrained by nuclear
	theory and experiment.}
	\label{fig:tditotcorr}
\end{figure}

As the value of the neutron star moment of inertia $I$ increases, so too does the statistical average  
of the neutron star radius. Our current EOS distribution results in $\langle R \rangle=12.01\,$km,
$R_{-\sigma}=11.36\,$km, 
$R_{+\sigma}=12.48\,$km, $R_{-2\sigma}=10.26\,$km, $R_{+2\sigma}=12.87\,$km
for a $1.4\,\msun$ neutron star, and the most probable radius is $\tilde{R}=12.15\,$km.
We apply Bayesian analysis to see how the credibility interval varies 
for a given moment of inertia measurement.
Bayesian statistics gives for the posterior probability
\begin{equation}
	P(\mathcal{M}_i | D) = \frac{P(D|\mathcal{M}_i)P(\mathcal{M}_i)}
	{\sum_j P(D|\mathcal{M}_j)P(\mathcal{M}_j) }\,,
\end{equation}
where $\mathcal{M}_i$ stands for the nuclear model parameters, $D$ represents the data set,
$P(\mathcal{M}|D)$ is the posterior probability,
$P(D|\mathcal{M})$ is the likelihood function,
and $P(\mathcal{M})$ is the prior distribution function, which in this case we take to be
that arising from the inclusion of EOS constraints from nuclear theory and experiment (that is, our
previous posterior distribution function). 
From $n$ measurements of the moment of inertia $I_k$ and the corresponding uncertainties $\sigma_k$, 
we define the likelihood function for a specific nuclear model as
\begin{equation}
	P(D|\mathcal{M}) = \prod_{k}^{n}\frac{1}{\sqrt{2\pi} \sigma_{k}}
	\mathrm{Exp} \left[ -\frac{(I(\mathcal{M}) - I_k)^2 }{2\sigma_{k}^2} \right] \,,
\end{equation}
where $I(\mathcal{M})$ is the moment of inertia from the specific nuclear model.

\begin{figure}[t]
	\centering
	\includegraphics[scale=0.45]{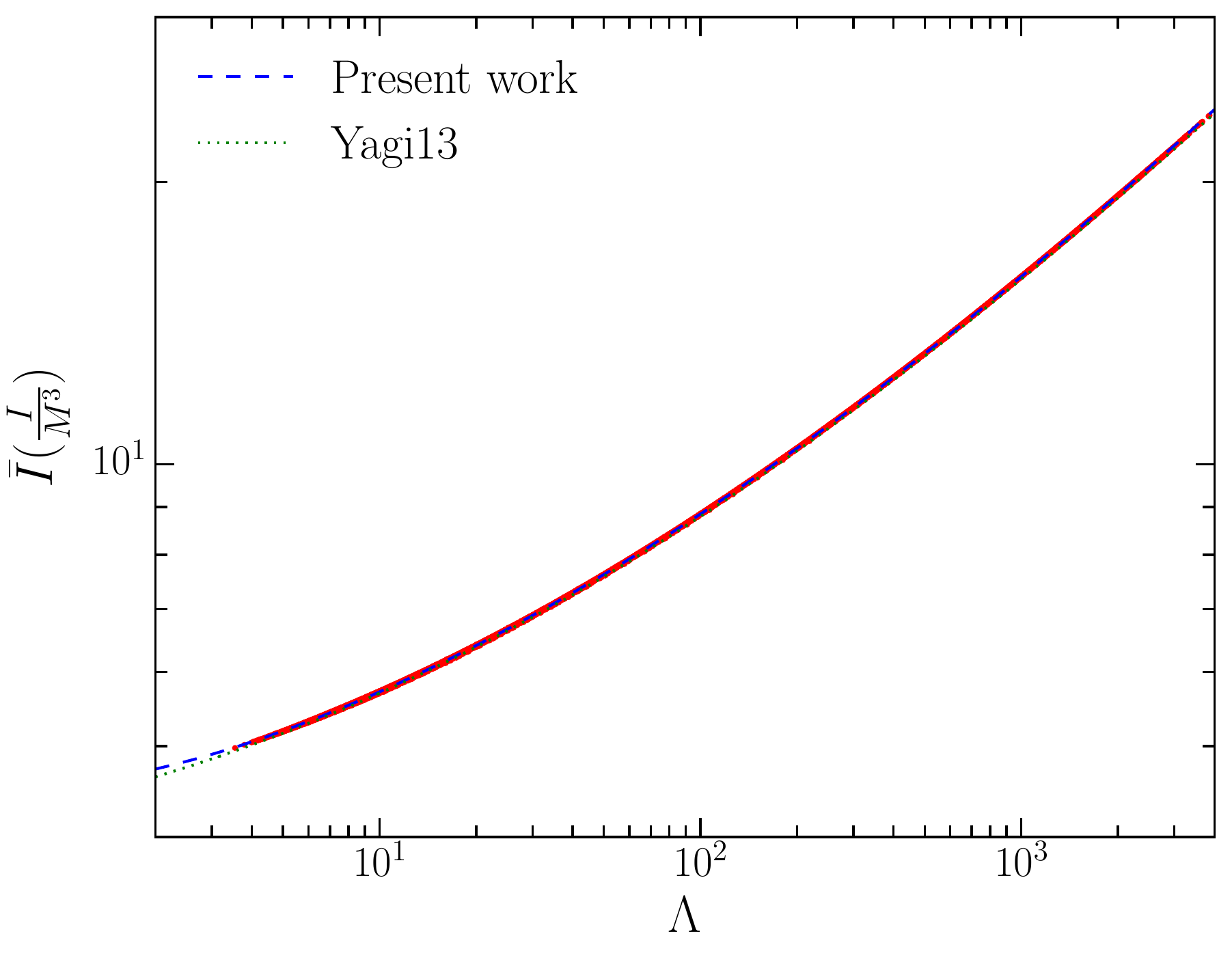}
	\caption{(Color online) Distribution (red) for $\bar{I} = I/M^3$ vs.\ the neutron star tidal deformability
	$\Lambda$. The blue-dashed line is the empirical relation Eq.\ (\ref{loglog}) derived in the present work, 
	and the green-dotted curve is that from Ref.\ \cite{yagi13}.}
	\label{fig:allmasstditot}
\end{figure}

In the top-right, bottom-left, and bottom-right panels of Fig.\ \ref{fig:bayemrdist} we show the resulting
posterior probability distribution for the neutron star mass-radius relation assuming the following moment
of inertia measurements for the $1.338\,\msun$ neutron star in PSR J0737-3039: 
$\{I_{45}^1=1.1\pm 0.11,\, I_{45}^2=1.3 \pm 0.13, \, I_{45}^3=1.5 \pm 0.15 \}$.
The resulting $1\sigma$ and $2\sigma$ credibility intervals for the radius of a $1.4\,\msun$ neutron star are shown in 
Table\,\ref{bayestab}.
\begin{table}[b]
	\begin{tabular}{cccccc}
		\hline
		\hline
		& $R_{-2\sigma}$\,(km) & $R_{-\sigma}$\,(km) & $\tilde{R}$\,(km) & $R_{+\sigma}$\,(km) & $R_{+2\sigma}$\,(km) \\ 
		\hline
		 -         & 10.26  & 11.36  & 12.15  & 12.48  & 12.87 \\
		$I_{45}^1$ &  9.97  & 10.64  & 11.35  & 11.76  & 12.20  \\ 
		$I_{45}^2$ & 10.89  & 11.46  & 12.05  & 12.35  & 12.71  \\ 
        $I_{45}^3$ & 11.33  & 11.82  & 12.30  & 12.61  & 12.93  \\ 
\hline
\end{tabular}
\caption{Bayesian analysis for the radius of a $1.4\,\msun$ neutron star from artificial moment of inertia
measurements at 10\% precision for three different cases.}
\label{bayestab}
\end{table}
For instance, in the case that $I_{45}^1=1.1 \pm 0.11$, the maximum radius for a $1.4\,\msun$ neutron
star would be shifted down to approximately $R_{\rm max} \simeq 12.2$\,km. Likewise, under the scenario
where $I_{45}^3=1.5 \pm 0.15$ the minimum radius would be shifted up to about $R_{\rm min} \simeq 11.3$\,km.
Even a moment of inertia measurement nearly consistent with our most probable value from the 
prior distribution will further constrain the equation of state. This is shown in the lower-left panel of
Fig.\ \ref{fig:bayemrdist}, where a measured value of $I_{45}^2=1.3 \pm 0.13$ increases the lower bound
on the radius of $1.4\,\msun$ neutron star up to roughly $10.9$\,km, even though the measured value 
of $I$ would be only a few percent below that of the most probable value from the prior distribution.

\begin{figure}[t]
	\centering
	\includegraphics[scale=0.45]{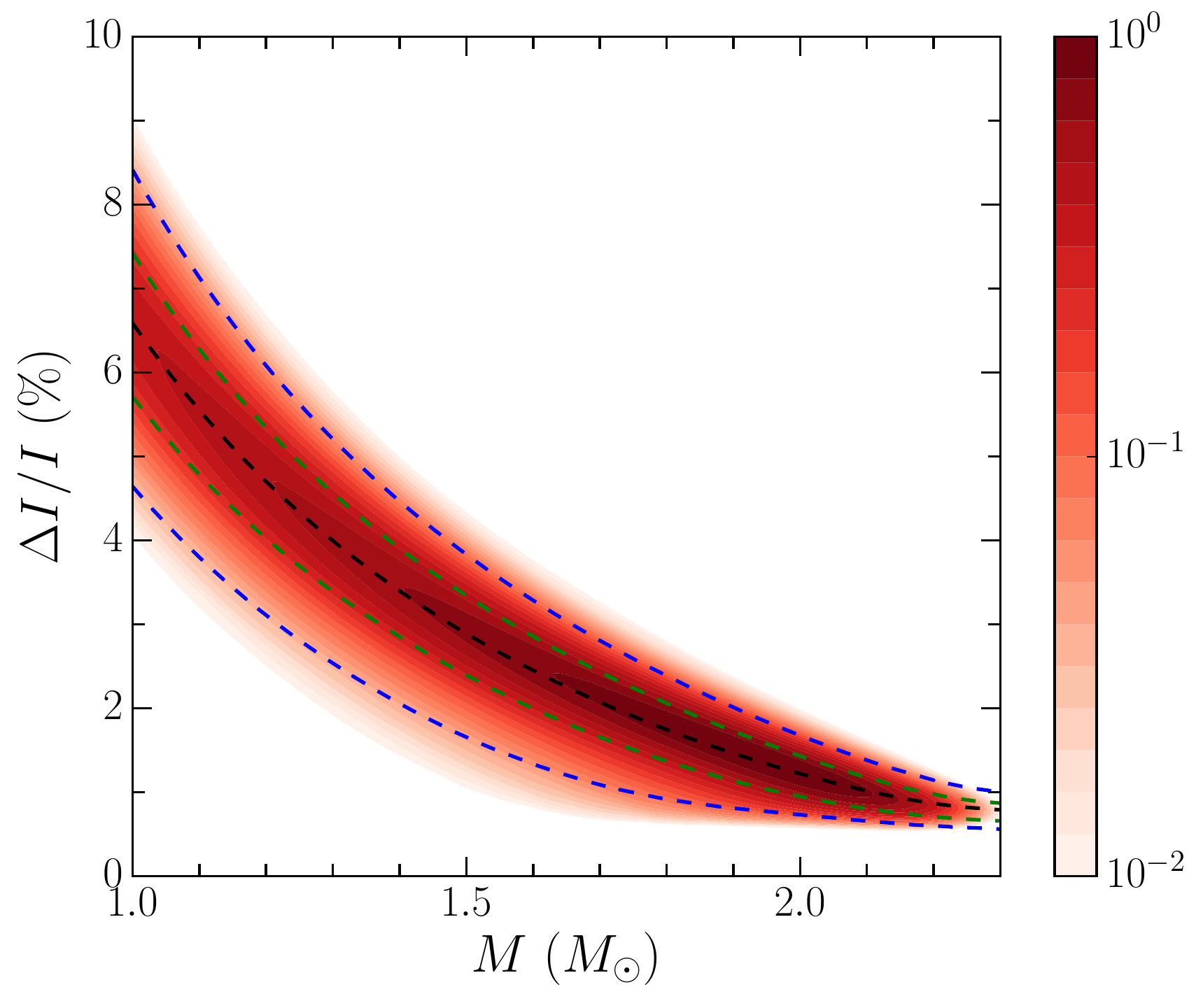}
	\caption{(Color online) Probability distribution for the ratio of the crust moment of inertia to the total
	 moment of inertia as a function of the neutron star mass. The most likely value is represented by the 
	 black dashed line, while the green- and blue-dashed lines represent the $1\sigma$ and $2\sigma$
	 credibility intervals, respectively.}
	\label{fig:iciratio}
\end{figure}

Even more strongly constraining is the relation between the neutron star moment of inertia
and dimensionless tidal deformability as pointed out in Ref.\ \cite{yagi13}. In 
Fig.\ \ref{fig:tditotcorr} we show the correlation between the tidal deformability and
the moment of inertia of a neutron star with mass $M=1.338\,\msun$.
We find the empirical formula,
\begin{equation}
	\Lambda = -37.109 + 181.540 ( I_{45} )^{3.5},\, ~\mathrm{for } \, M=1.338\,\msun\,.
\end{equation}
Note that this formula may support different values for the coefficients when neutron stars with different
masses are considered, due to the fact that the most probable values of $I$ and $\Lambda$ vary with the 
mass. As seen in Fig.\ \ref{fig:tditotcorr} precise measurements of neutron star moments of inertia can be 
used to very tightly constrain the tidal deformability, and vice versa.

In Fig.\ \ref{fig:allmasstditot} we demonstrate the strong correlation between $\bar{I}=I/M^3$ and 
$\Lambda$ for all of the neutron stars that may be constructed from our Bayesian posterior probability 
distribution for the nuclear equation of state. The red solid line in Fig.\ \ref{fig:allmasstditot} 
results from plotting individually the 
scaled moments of inertia vs.\ tidal deformabilities $\Lambda$.
The log-log functional relationship between $\bar{I}$ and 
$\Lambda$ can be well approximated by
\begin{equation}
	\mathrm{Log}_{10}( \bar{I} ) = 0.65974 + 0.097374 \left[ \mathrm{Log}_{10} (\Lambda)\right]^{1.56}\,.
	\label{loglog}
\end{equation}
This is shown as the blue dashed line in Fig.\ \ref{fig:allmasstditot}. In comparison we also show as the 
green dashed line in Fig.\ \ref{fig:allmasstditot}, the original correlation derived in Ref.\ \cite{yagi13}. 
While both functions fit the theoretical results very well, we find that our functional form has a smaller 
$\chi^2$ value with fewer parameters. 

\begin{figure}[t]
	\centering
	\includegraphics[scale=0.4]{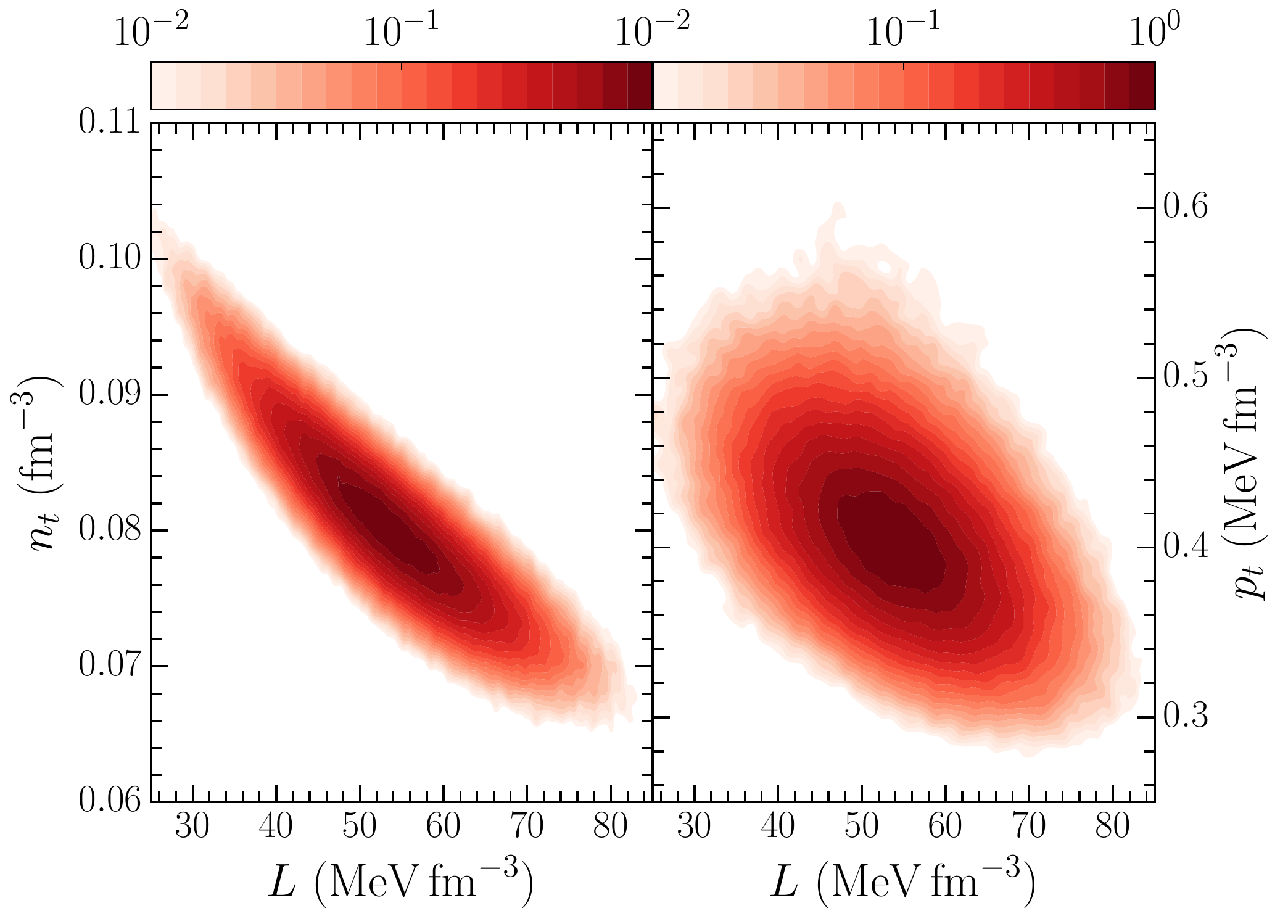}
	\caption{(Color online) Contour plot of transition density and transition pressure at the core-crust boundary
	as a function of the symmetry energy slope parameter $L$ at nuclear saturation density.}
	\label{fig:npcrustlv}
\end{figure}

Finally, we investigate the fraction of the neutron star moment of inertia contained in the crust.
This quantity is related to the ratio of the superfluid angular momentum to the total angular momentum
in the neutron star, which must be sufficiently large in order to support the observed glitch activity of 
the Vela pulsar.
One can derive \cite{link99} that in the standard hydrodynamic two-fluid model, the ratio of the 
crustal moment of inertia $\Delta I$ to the total moment of inertia $I$ must satisfy
\begin{equation}
\frac{\Delta I}{I} \ge 0.014.
\label{c1}
\end{equation}

However, strong entrainment of otherwise free neutrons by the inner crust can be studied within
band theory calculations and has been shown \cite{chamel12,andersson12,chamel13} 
to reduce the neutron superfluid angular momentum reservoir. 
The key quantity is the neutron effective mass, defined as\,\cite{chamel13}
\begin{equation}
	\begin{aligned}
		m_n^* \equiv m_n n_n^f/n_n^c,
	\end{aligned}
\end{equation}
where $m_n$ is the bare neutron mass, $n_n^c$ is the density of conduction neutrons, and
$n_n^f$ is the density of unbound neutrons. The neutron effective mass strongly depends on
the density and peaks at a value of $m_n^* \simeq 10\, m_n$ around $n=0.025$\,fm$^{-3}$. Averaging 
over typical densities in the crust leads to a decrease in the superfluid angular momentum reservoir such
that $\Delta I/I \gtrsim 0.07$ is required to explain observed glitch activity. Such large crustal moments
of inertia are not favored in most theoretical modeling of neutron star structure, especially for
the moderately soft equations of state produced by chiral effective field theory. 

\begin{figure*}[t]
	\centering
	\includegraphics[scale=0.7]{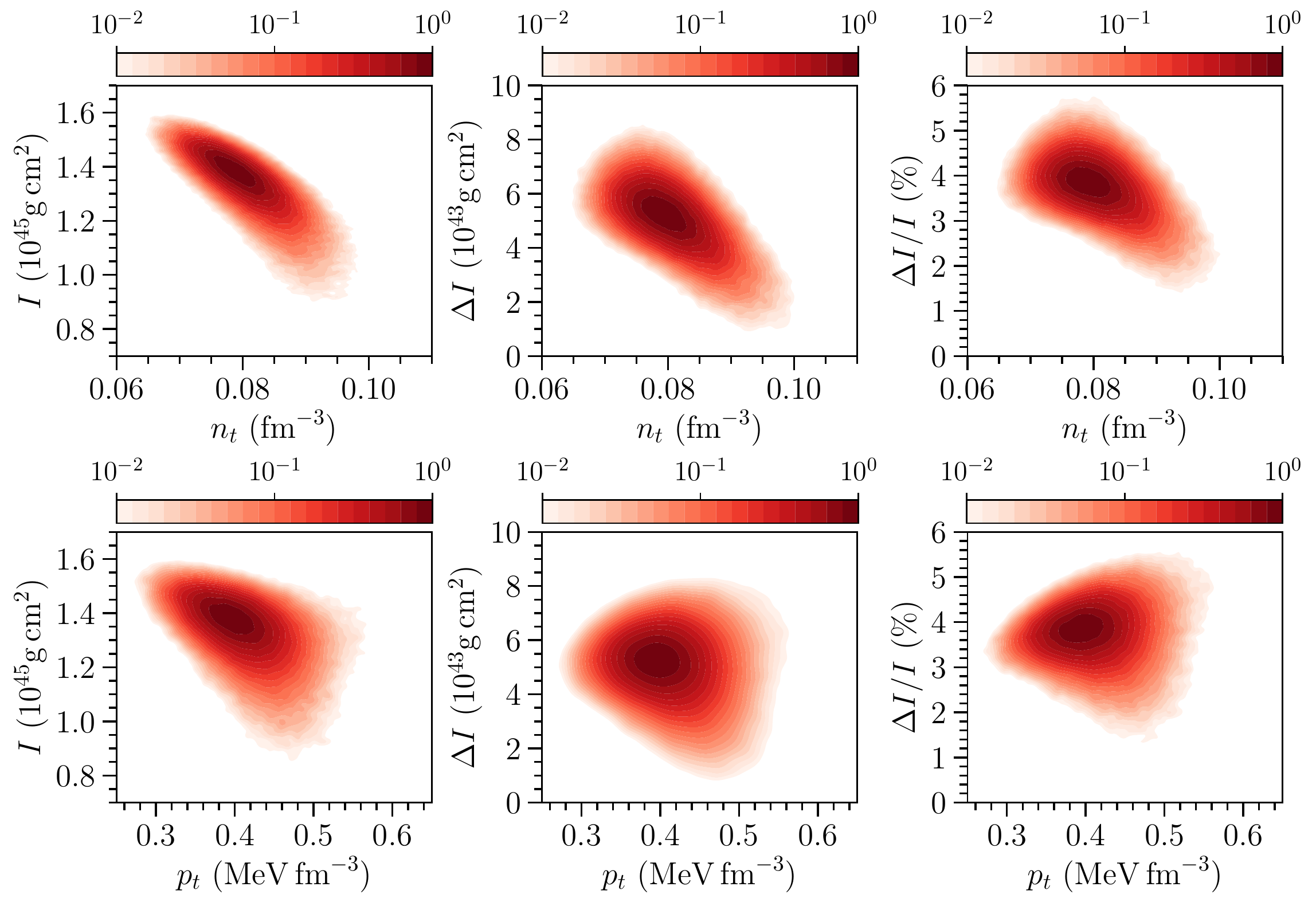}
	\caption{(Color online) Contour plots for the total moment of inertia $I$ (left), crust moment of inertia
	$\Delta I$ (middle) and fractional crustal moment of inertia $\Delta I/I$ (right) vs. the transition density 
	(top) and transition pressure (bottom) for a $M=1.338\,\msun$ neutron star.}
	\label{fig:lvpn0p2n0m1338}
\end{figure*}

In Fig.\ \ref{fig:iciratio} we show the fraction of the crustal moment of inertia to the total moment of inertia
as a function of the neutron star mass. The crust moment of inertia is obtained numerically from the Eq.\,\eqref{eq:icrust}. The blue (green) dashed curves in Fig.\ \ref{fig:iciratio} indicate 
$\pm 2\sigma$ ($\pm 1\sigma$) credibilities, while
the black dashed curve represents the central value of the momentum fraction.
Low-mass neutron stars tend to have thicker crusts than
high-mass neutron stars, leading to fractional crustal moments of inertia that are systematically larger.
Although the mass of the Vela pulsar is not precisely known, recent work \cite{ho15} has estimated
a value $M = 1.51 \pm 0.04\,\msun$. From Fig.\ \ref{fig:iciratio} we see that the present EOS modeling
within the scenario of strong neutron entrainment would be insufficient to explain large pulsar glitch
activity. However, such large entrainment effects have
recently been called into question, and in particular the inclusion of neutron pairing in band theory
calculations may significantly reduce neutron entrainment and increase the angular momentum
reservoir \cite{watanabe17}. In our models, $\Delta I / I \simeq 1.8-3.9\%$ for a $1.5\,\msun$ neutron star, 
and therefore a reduction in the average neutron entrainment in the crust from $n_n^f / n_n^c \simeq 5$ 
in the scenario of strong entrainment to $n_n^f / n_n^c \rightarrow 1.0-2.8$ would be sufficient 
to explain the glitch activity of the Vela pulsar.

Connecting the crustal fraction of the moment of inertia to specific features of the nuclear
equation of state is challenging. It has been suggested \cite{link99,fattoyev10} that the pressure 
of beta-equilibrium matter at the neutron star core-crust interface is strongly correlated with the 
crustal moment of inertia:
\begin{equation}
	\Delta I \sim R_t^6 \,p_t\,,
\label{rpt}
\end{equation}
where $R_t$ is the radius of the neutron star core and $p_t$ is the pressure at the core-crust boundary.
Eq.\ (\ref{rpt}) can be derived under the approximation of slow rotation and thin, low-density crusts.
In Fig.\ \ref{fig:npcrustlv} we plot the probability contour plot of the crustal 
density $n_t$ and
corresponding pressure $p_t$ at the core-crust interface as a function of the nuclear symmetry energy slope
parameter $L = 3n_0 \frac{\partial E_{\rm sym}}{ \partial n}|_{n_0}$.
We note here the negative correlation between $n_t$ and $L$, as well as
a weaker negative correlation between $p_t$ and $L$.
This negative correlation results in a related negative correlation between $n_t$ and $\Delta I/I$
and almost no correlation between $p_t$ and $\Delta I/I$ as shown in the right panels of
Fig.\ \ref{fig:lvpn0p2n0m1338}, plotted for a neutron star with mass $M=1.338\msun$.
That is, a low transition density at the core crust boundary 
results in the crustal moment of inertia making up a higher fraction of the total.

In the left and center panels of Fig.\ \ref{fig:lvpn0p2n0m1338} we show also the probability 
distributions for the total moment of inertia and the crustal moment of inertia
as a function of the transition density and pressure for a $1.338\msun$ neutron star. 
We find negative correlations with
the transition density and relatively weak correlations with the transition pressure. Physically,
as the transition density increases, $L$ decreases and the neutron star becomes more compact.
Thus the moment of inertia decreases. On the other hand, $p_t$ is anti-correlated with $I$ since 
$p_t$ is anti-correlated with $L$ and $L$ correlates with $R$, the radius of the neutron star. 

\begin{figure}[t]
	\centering
	\includegraphics[scale=0.52]{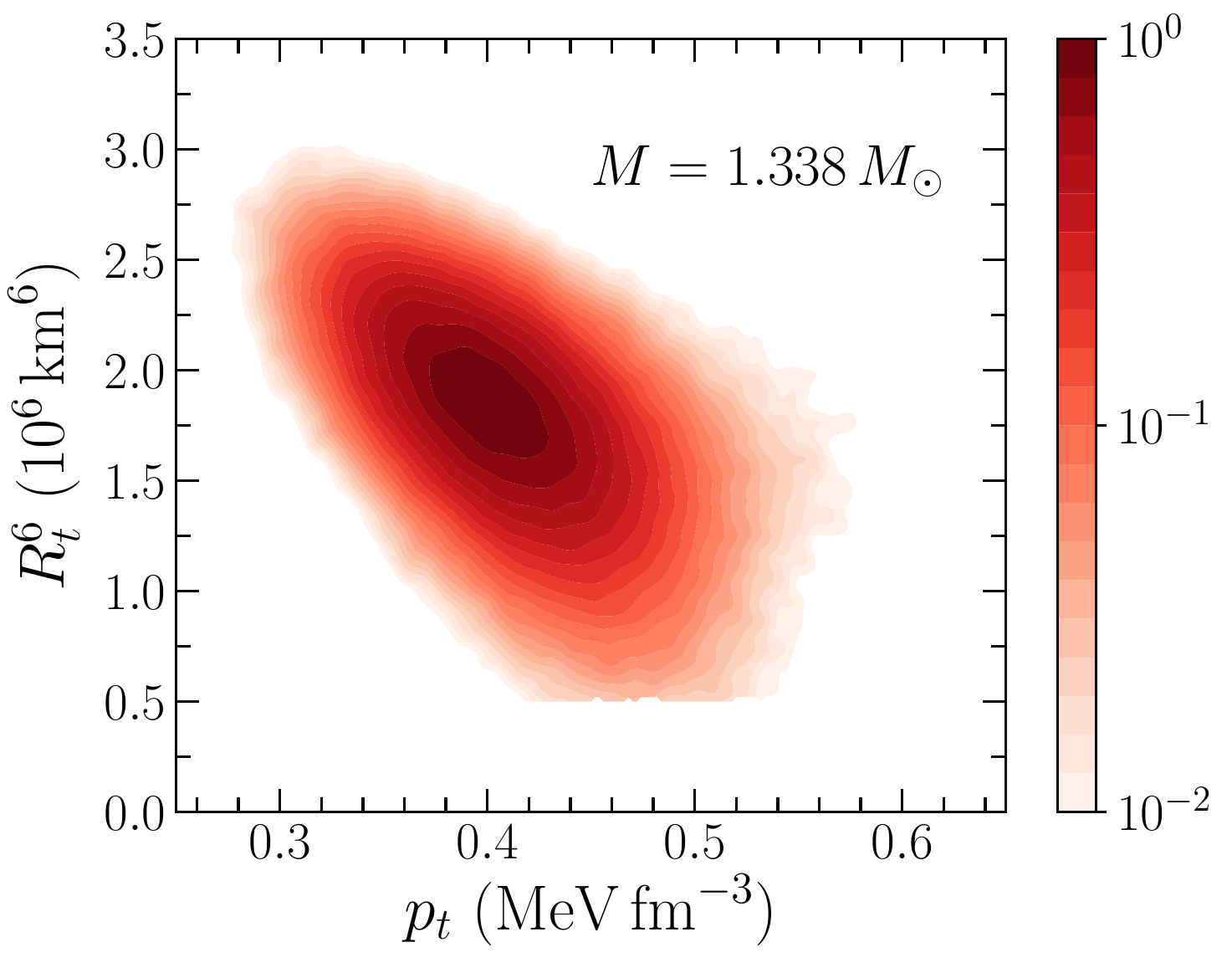}
	\caption{(Color online) Contour plot of the neutron star core radius $R_t^6$ vs.\ the crust-core 
	transition pressure $p_t$ for a $1.338\,\msun$ neutron star. We consider 300,000 equations of state
	generated from Bayesian modeling of the nuclear energy density functional.}
	\label{fig:rptcrust}
\end{figure}

Since we find no strong link between the crustal moment of inertia and the core-crust
transition pressure (in contrast to previous works), we examine in more detail the 
correlation between the transition pressure
$p_t$ and the core radius $R_t$, which was not considered in Refs.\ \cite{link99,fattoyev10}.
In Fig.\ \ref{fig:rptcrust} we plot the probability distribution of the core radius $R_t^6$ vs.\ the
core transition pressure. We find a statistically significant anticorrelation between the two quantities,
which from Eq.\ (\ref{rpt}) reduces the dependence of the crustal moment of inertia on the 
transition pressure. Indeed soft equations of state (with low values of $L$) are correlated with 
higher transition pressures as seen in Fig.\ \ref{fig:npcrustlv} and also give rise to more
compact neutron stars with smaller core radii. The combined result is almost no correlation
between the crustal moment of inertia and the transition pressure, as seen in the middle panel
of Fig.\ \ref{fig:lvpn0p2n0m1338}.

\section{Summary}\label{sec:sum}
We have computed the moment of inertia of neutron stars based on a Bayesian analysis
of the nuclear energy density functional constrained by chiral effective field theory
and nuclear matter properties deduced from finite nuclei. We predict that for pulsar 
PSR J0737-3039A, with a well measured mass of $M = 1.338\msun$, the 
moment of inertial lies in the range 
$1.04 \times 10^{45}$\,g\,cm$^{2} < I < 1.51 \times 10^{45}$\,g\,cm$^{2}$ at the 95\% credibility
level, while the most probable value for the moment of inertia is 
$\tilde I = 1.36 \times 10^{45}$\,g\,cm$^{2}$. We have also shown that a pulsar timing measurement 
of the PSR J0737-3039A moment of inertia to 10\% precision will result in meaningful constraints 
on the current Bayesian modeling of the equation of state by imposing the likelihood function
for the posterior probability. Three scenarios were considered $I_{45} = 
\{1.1, 1.3, 1.5\}$ and the resulting effect on the neutron star mass-radius relation were analyzed.
In particular, we find that the credibility interval for the radius of a $1.4\,\msun$
neutron star decreases from $R_{+2\sigma} - R_{-2\sigma}=2.6\,$km to $2.6$, $1.8$ and $1.6$\,km 
depending on the moment of inertia measurement.

We have studied as well correlations among the neutron star moment of inertia, radius, and tidal
deformability. A strong correlation is demonstrated between the moment of inertia and tidal deformability, 
indicating that one of the two quantities will strongly constrain the other. From our large sample of 
realistic equations of state, we derived a new functional model for the $I/M^3$ vs. $\Lambda$
universal relation.

Finally, we have employed realistic modeling of the crust equation of state to determine the 
fraction of the neutron star moment of inertia contained in the crust. We find that for typical
neutron star masses of 1.2-1.5\,$M_\odot$, the fractional crustal moment of inertia is less than $7\%$.
In the strong neutron entrainment scenario, our small values of the crustal moment of inertia would be 
unable to account for the observed large glitch activity in the Vela pulsar. We have also shown that
the crustal moment of inertia is weakly correlated with the core-crust transition density. 
Low transition densities at the core-crust boundary allow a large ratio between the moment of inertia 
of the neutron star crust to the total moment of inertia, since the high transition pressure is responsible 
for smaller core radii in neutron stars. In contrast to previous works, we find no correlation between 
the crustal moment of inertia and the transition pressure. It is understood that the symmetry energy 
slope parameter $L$ is anti-correlated with $p_t$ but positively correlated with $R$.
Thus, their effects counteract each other, and no visible correlation is seen.

In the future we plan to incorporate as well neutron star radius measurements from the NICER mission and 
neutron star tidal deformability measurements the LIGO/VIRGO collaborations within the present Bayesian 
statistical modeling of the equation of state.



\acknowledgments

Work supported by the National Science Foundation under Grant No.\ PHY1652199. 
Portions of this research were conducted with the advanced computing resources provided by 
Texas A\&M High Performance Research Computing.

\bibliographystyle{apsrev4-1}

\end{document}